\documentclass[twocolumn,amsmath,amssymb,floatfix,pra,aps]{revtex4}

\usepackage{graphicx}% Include figure files

\def\nm{{\ {\rm nm}}}                       % nm
\def\mm{{\ {\rm mm}}}                       % mm
\def\cm{{\ {\rm cm}}}                       % cm
\def\meter{{\ {\rm m}}}                     % cm
\def\micron{{\ \mu{\rm m}}}                 % microns
\def\g{{\ {\rm g}}}                     % Gram

\def\liter{{\ {\rm L}}}                     % liter

\def\mbar{{\ {\rm mbar}}}                       % Torr
\def\tesla{{\ {\rm T}}}                     % tesla
\def\mT{{\ {\rm mT}}}                       % mT
\def\uT{{\ {\rm \mu T}}}                        % uT
\def\amp{{\ {\rm A}}}                       % Amp

\def\kW{{\ {\rm kW}}}                       % W
\def\watt{{\ {\rm W}}}                      % W
\def\mW{{\ {\rm mW}}}                       % mW
\def\uW{{\ \mu{\rm W}}}                 % uW
\def\Hz{{\ {\rm Hz}}}                       % Hz
\def\MHz{{\ {\rm MHz}}}                     % MHz
\def\ms{{\ {\rm ms}}}                       % ms
\def\second{{\ {\rm s}}}                    % s

\def\C{{\ {\rm C}}}                 % K
\def\kelvin{{\ {\rm K}}}                    % K
\def\uK{{\ \mu{\rm K}}}                     % uK
\def\Rb87{^{87}\rm{Rb}}                 % Rb 87
\def\Li6{^{6}\rm{Li}}                   % Li 6
\newcommand{\ket}[1]{|#1\rangle}

\begin{document}

\title{Rapid production of $^{87}$Rb BECs in a combined magnetic and optical potential}

\author{Y.-J.~Lin}
\author{A.~R.~Perry}
\author{R.~L.~Compton}
\author{I.~B.~Spielman}
\email{ian.spielman@nist.gov}
\author{J.~V.~Porto}
\email{trey@nist.gov}
\affiliation{Joint Quantum Institute, National Institute of Standards and Technology, and University of Maryland, Gaithersburg, Maryland, 20899, USA}

\date{\today}

\begin{abstract}
We describe an apparatus for quickly and simply producing $\Rb87$
Bose-Einstein condensates.  It is based on a magnetic quadrupole
trap and a red detuned optical dipole trap.  We collect
atoms in a magneto-optical trap (MOT) and then
capture the atom in a magnetic quadrupole trap and force rf
evaporation.  We then transfer the resulting cold,
dense cloud into a spatially mode-matched optical dipole trap by
lowering the quadrupole field gradient to below gravity.  This
technique combines the efficient capture of atoms from a MOT into a
magnetic trap with the rapid evaporation of optical dipole traps;
the approach is insensitive to the peak quadrupole gradient and the
precise trapping beam waist. Our system reliably produces a condensate with
$N\approx2\times10^6$ atoms every $16\second$.
\end{abstract}
%\pacs{}

\maketitle

A range of techniques have been developed to produce quantum
degenerate gases, of varying degrees of complexity and difficulty.
Almost all current methods rely on the same basic approach: laser
cooling of atoms~\footnote{Bose-condensation of atomic hydrogen is
an exception where the initial cold atomic sample was prepared not
by laser cooling, but instead by thermalization with a buffer gas
~\cite{Fried1998}.}, followed by evaporative cooling in a
conservative trap~\cite{Anderson1995,Davis1995a}. When designing
there are many often competing considerations: reliability, speed,
simplicity, large optical access, and reasonably large number.  We
describe here a simple approach that quickly produces a relatively
large Bose-Einstein condensate (BEC) of $\Rb87$ atoms.

Magnetic quadrupole traps for neutral atoms~\cite{migdall85a} have
several strengths. They have large trap volumes, which can be well
matched to the size of laser cooled atom clouds. The effectively
linear potential provides tight confinement, allowing for efficient
evaporative cooling, and the quadrupole field can be generated from
a simple arrangement of two electromagnetic coils which allows for
good optical access to the sample.  The quadrupole trap's one major
drawback is Majorana spin-flip losses and the resulting heating near
the zero-field point at the center of the trap. This limits forced
evaporative cooling to relatively low phase-space
densities~\cite{Davis1995}.  Still, the first two dilute-gas BEC's
were produced in traps derived from the simple quadrupole trap,
taking advantage of its strengths by avoiding the Majorana losses in
two different ways: by providing a time-orbiting bias field (``TOP''
trap) to shift the zero away from the atom cloud~\cite{petrich95a,
Anderson1995}, or by using a repulsive optical potential (i.e., an
optical plug) to push the atoms away from the
zero~\cite{Davis1995a}. The TOP trap is still widely used, and the
optically plugged trap has successfully been
revisited~\cite{naik05blochraizen}.

Far detuned optical dipole traps have a different set of strengths.
They can be spin-state independent, made in flexible geometries that
provide good optical access, have tight confinement and efficient
evaporation, and they don't require magnetic coils. One drawback is
that simultaneously deep and large-volume traps require
prohibitively  large laser power.  One must carefully ``mode match''
the laser-cooled atoms to the optical dipole trap, and all-optical
approaches tend to produce relatively small
BEC's~\cite{barrett01a,weber03a,cennini03a}.  By loading into
large-volume, shallow traps and transferring into smaller volume
traps~\cite{weber03a}, larger BEC's can be
made~\cite{Kinoshita2005}, but this requires superlative
high-density laser cooling.  It has been suggested that BEC's could
be efficiently produced using a quadrupole trap as a reservoir to
directly feed a red-detuned dipole trap~\cite{comparat06a}.  We
demonstrate here such a hybrid technique which combines the
advantages of quadrupole and optical traps while avoiding their
individual weaknesses. We can then load the BEC from the hybrid trap into an
all-optical dipole trap without an additional blue-detuned optical plug.

The basic approach is to load laser cooled atoms into a quadrupole
trap, and use forced rf evaporation~\cite{Davis1995} to reduce the
phase space density until Majorana spin-flips cause significant loss
and heating. We then transfer atoms from the quadrupole trap to a
single-beam optical dipole trap. This step can be fairly efficient,
transferring a large fraction of the atoms from the quadrupole trap
into the optical trap. To limit Majorana losses during and after the
transfer, the center of the optical trap is offset by roughly a beam
waist from the field zero of the magnetic trap.  Further, offset
quadrupole field provides harmonic confinement along the beam
direction; this approach therefore yields 3D confinement without a crossed
dipole trap. Since the atom cloud is adiabatically cooled during the
transfer, only a relatively small optical trap depth ($\sim
49$~$\mu$K in our case) is required. The change of the trap shape
during the adiabatic transfer, from linear quadrupole to harmonic,
modifies the density of states in such a way as to increase the
phase space density at constant entropy, leading to fairly high
initial phase space density in the dipole trap. Forced evaporative
cooling by lowering the optical trap power~\cite{barrett01a} then
leads to quantum degeneracy.  This scheme is simple, flexible and
allows for significant optical access. We reliably produce $^{87}$Rb
BEC's with $N=2\times10^6$ atoms in a time $t\approx16\second$.

\section{Theory}

In order to discuss this approach, we first provide a brief overview
of trapped gas thermodynamics and cooling for our trap geometry.
(More detailed discussions can be found in
Refs.~\cite{luiten96a,ketterle96b} and further references therein.)
The thermodynamic properties of the gas are determined by the
partition function $\zeta = V_0/\Lambda^{3}$, where $\Lambda=(2 \pi
\hbar^2 / m k_B T)^{1/2}$ is the thermal de Broglie wavelength and
\begin{equation}
V_0  =  \int \mathrm{exp}[-U({\bf r})/k_B T] \ d^3r \label{eq:V0}
\end{equation}
is an effective trap volume, where $U({\bf r})$ is the trapping
potential with the energy minimum $U({\bf r_{\rm min}})=0$ at
position $\bf r_{\rm min}$, and $T$ is the temperature. Corrections to $V_0$ due to
the finite trap depth $\epsilon_t = \eta k_B T$ during evaporation
can be treated using a truncated Boltzmann
distribution~\cite{luiten96a}. The large-$\eta$ limit,
Eqn.~\ref{eq:V0}, is reasonable for most purposes when $\eta \gtrsim
8$, and below this range provides a qualitative description. From
$\zeta$ and $V_0$, one can calculate, for example, the free energy
$A = -N k_B T \ln{\zeta} $, the entropy $S = -\partial A/\partial
T$, the peak phase space density $D=N/\zeta$ and the density
distribution
\begin{equation}
n({\bf r}) =n_0\ \mathrm{exp}[{-U({\bf r})/k_B T}]=\frac{N}{ V_0}
\mathrm{exp}[{-U({\bf r})/k_B T}].
\end{equation}
Dynamic rates relevant for evaporation, such as $n$-body loss,
collision rates, evaporation rates, etc., can be calculated from
similar integrals~\cite{luiten96a}.

\begin{figure}[t]
\begin{center}
\includegraphics[width=3.5in]{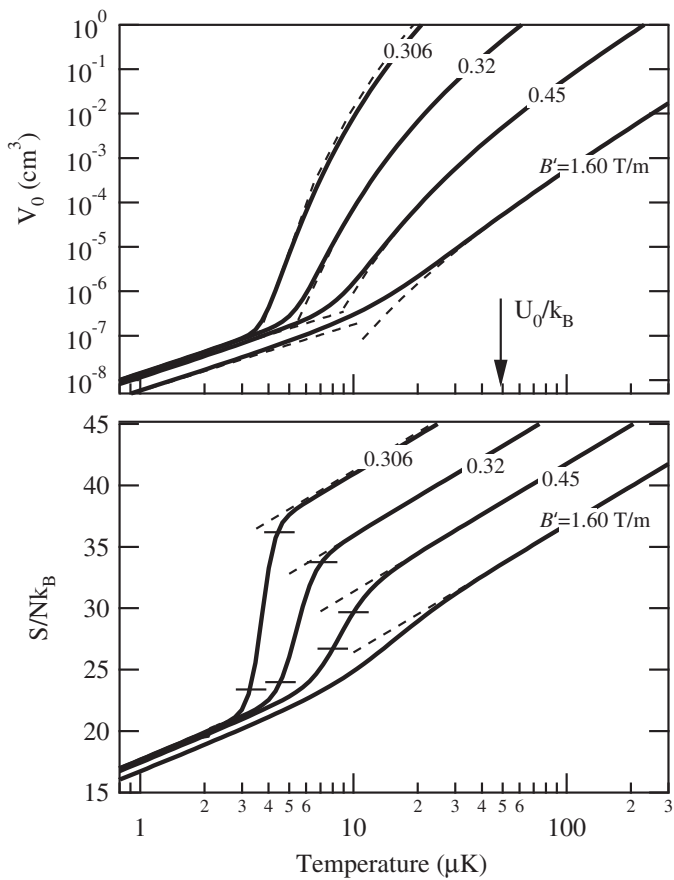}
\end{center}
\caption{{\em Top:} $V_0$ as a function of temperature for several
different field gradients. Thick lines: numerical calculation for
field gradients  $B^\prime = 1.60$, 0.45,  0.32 and 0.306~T/m. (For $\Rb87$ magnetically trapped in $|F,m_F\rangle=|1,-1\rangle$, $m g=\mu \times 0.30492$~T/m.) Dashed
lines: analytical approximations, Eqns.~\ref{eq:V0high}
and~\ref{eq:V0low}. The dipole trap depth for all plots (indicated
by the arrow) is $U_0/k_B=49$~$\mu$K, and the beam waist is
65~$\mu$m along $\hat x$ and 78~$\mu$m along $\hat z$ (which accounts for the
difference in measured trap frequencies along $\hat x$ and $\hat z$). {\em
Bottom:} The entropy per particle, $S(T)/N k_B$, for the same field
gradients in the dipole-plus-quadrupole trap (solid lines) and for
quadrupole only traps (dashed lines). The short horizontal marks
crossing the curves indicate the regions where the atoms are
transferred into the dipole trap: the upper set marks where 10\% of
the atoms are in the dipole trap, and the lower set marks where 90\%
of the atoms are in the dipole trap.  } \label{fig:V0andS}
\end{figure}

The effective potential for atoms in the combined quadrupole plus dipole trap, including gravity is
\begin{widetext}
\begin{equation}
U({\bf r}) = \mu B^\prime \sqrt{\frac{x^2}{4}+\frac{y^2}{4}+z^2} -
U_0 \mathrm{exp}\{{-2
   \left[x^2+(z-z_0)^2\right]/w_0^2}\}  + m g z +E_0.
\end{equation}
\end{widetext}
where $B^\prime$ is the quadrupole field gradient along $\hat{z}$,
$U_0, w_0$ and $z_0$ are the dipole beam trap depth, waist and
offset from the zero field point at $x,y,z=0$. $E_0$ is the energy
difference between the zero field point absent the dipole trap and the
total trap minimum, giving the trap minimum $U({\bf r_{\rm min}})=0$. $\mu$ and
$m$ are the magnetic moment and mass of the atom, respectively, and $g$ is the
acceleration due to gravity. Here the dipole beam is aligned along
$\hat y$, and is displaced vertically (along $\hat{z}$) below the
magnetic field zero. (We have
ignored the focusing of the beam along $\hat{y}$, since for our
waist $w_0=65~\micron$ the Raleigh length $\pi w_0^2 / \lambda
\approx 9\mm$ is large.) The effective volume as a function of
temperature can be calculated from Eqn.~\ref{eq:V0}.

At high temperature, by approximating $U({\bf r})$ as
\begin{equation}
U({\bf r}) \simeq  \mu B^\prime \sqrt{\frac{x^2}{4}+\frac{y^2}{4}+z^2} + E_0 + m g z,
\end{equation}
we have
\begin{equation}
V_0(T)  \simeq
   \frac{ 32 \pi e^{-E_0/k_B T}}{\left[ 1- (m g/ \mu B^\prime)^2\right]^2} \left(\frac{k_B T}{\mu B^\prime}\right)^3,
   \ \ {\rm for}\ \ T \gg U_0/k_B ,\label{eq:V0high}
\end{equation}
where $E_0 \simeq U_0$ for a typical dipole beam offset $z_0 \simeq w_0$.
 This approximation is only valid for field
gradients that compensate gravity, $\mu B^\prime > m g $. The
thermodynamic effect of gravity can be viewed as re-scaling the
field gradient to $\mu B^\prime_{eff} = \mu B^\prime [1- (m g/\mu
B^\prime)^2]^{2/3}$, which vanishes at $\mu B^\prime = mg$. In the
absence of the dipole trap and ignoring gravity, the effective
volume reduces to the simple quadrupole form $V_0(T) = 32 \pi (k_B
T/\mu B^\prime)^3$, but for most parameters we encounter, both
corrections are important.

\begin{figure}[t]
\begin{center}
\includegraphics[width=3.5in]{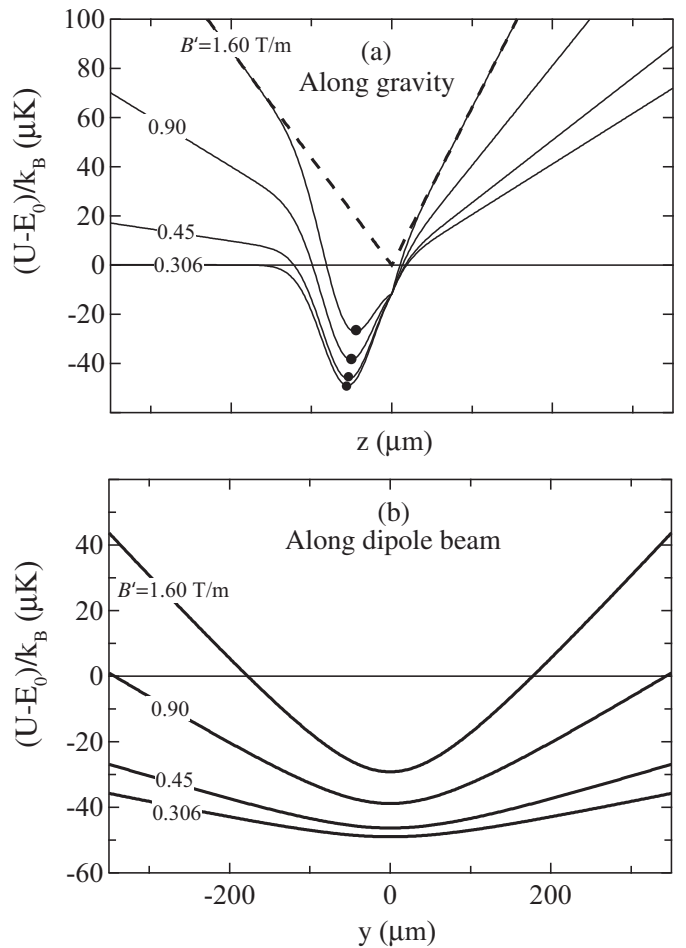}
\end{center}
\caption{Cross-sections of trapping potential with the offset $E_0$ subtracted, $U({\bf r})-E_0$,
at several points during an adiabatic expansion
($B^\prime = 1.60$, 0.90, 0.45, and 0.306~T/m): (a) Along $\hat z$ (gravity is along $-\hat z$) at $x,y=0$,
 the potential minima $z = z_{\rm min}$ are indicated by filled
circles, and (b) Along $\hat y$ (the dipole beam direction) at $x=0$, each taken at $z = z_{\rm min}$. The
dashed line is the potential at 1.60~T/m without the dipole trap.
The dipole trap parameters are the same as for
Fig.~\ref{fig:V0andS}.  } \label{fig:Uexpansion}
\end{figure}

At low temperature the dipole potential can be approximated as a harmonic trap
\begin{equation}
U({\bf r}) \simeq \frac{1}{2}\left[U^{\prime \prime}_x x^2+U^{\prime
\prime}_y y^2+U^{\prime \prime}_z (z-z_{\rm min})^2 \right],
\end{equation}
where $U_i^{\prime \prime}$ are the curvatures at the bottom of the
total trap, and the trap minimum is at position $z_{\rm min}$.
This gives
 \begin{equation}
   V_0(T)  \simeq  \frac{\ (2\pi k_B T)^{3/2}}{\sqrt{U^{\prime \prime}_x  U^{\prime \prime}_y U^{\prime \prime}_z}},\ \ {\rm for}\ \ T\ll U_0/k_B. \label{eq:V0low}
 \end{equation}
The curvatures along $\hat x$ and $\hat z$ are dominated by the
dipole trap, $\omega_{x,z} \simeq 2 \sqrt{U/m w_0^2}$, and the
curvature along $\hat y$ is dominated by the magnetic trap,
$\omega_x = (1/2)\sqrt{\mu B^\prime/m z_{\rm min}}$. As shown in
Fig.~\ref{fig:V0andS}, both the high and low temperature
approximations for the volume $V_0$ and entropy $S$ are quite good over a range of temperatures.
The high temperature approximation is accurate well below $U_0/k_B$.
%(for very near $\mu B^\prime \simeq m g$, where the numerical integration
%of Eqn.~\ref{eq:V0} has difficulty converging due to the large
%$V_0$).

The success of our hybrid approach depends critically on managing two processes: limiting Majorana
loss in the quadrupole trap and effective adiabatic (approximately constant entropy)
transfer from the quadrupole trap to the harmonic dipole trap.
Majorana loss is particularly detrimental, as it causes both loss
and heating. Such loss can be difficult to describe accurately, but
a simple argument~\cite{petrich95a} leads to the estimate that the
Majorana loss rate $\Gamma_m$ scales as $\Gamma_m \propto \hbar/ m
l^2$, where $l$ is the radial half width half maximum cloud size in
the quadrupole trap.  Using the proportionality constant measured
in~\cite{petrich95a}, we can estimate
\begin{eqnarray}
\Gamma_m & =&  3.6 \frac{ \hbar}{m l^2}\nonumber \\
 & = & 1.85
 \frac{ \hbar}{m}\left(\frac{\mu B^\prime}{k_B T}\right)^2,
 \label{eq:Majorana}
 \end{eqnarray}
where we have ignored gravity. This estimate is independent of the
elastic collision rate in the trap since it assumes thermal
equilibrium, an invalid assumption at low densities or large loss
rates. Nonetheless, it suggests that for field gradients of $\sim
1.40$~T/m, the lifetime should be on the order of one second at
20~$\mu$K. Near this temperature fast evaporation (high elastic
collision rate) is advantageous. This loss rate can be mediated
somewhat by adiabatically expanding the trap, since the temperature
scales more weakly than the field gradient at constant entropy, $T
\propto (B^\prime)^{2/3}$, and weakening the trap will decrease the
loss rate, at the expense of a lower collisional rate.

The adiabatic transfer is more subtle than a mere expansion of the
trap volume, since in addition to expanding the trap, the shape of
the trap is modified. This leads to changes in phase-space density,
even at constant entropy~\cite{stamper-kurn98a}, an effect used to
obtain a BEC of cesium from a very cold, but dilute
gas~\cite{weber03a}. In our case, the effect is quite strong, as the
transfer process both decreases the temperature and increases phase
space density.

The trap potential during an example expansion sequence is shown in
Fig.~\ref{fig:Uexpansion}. The change of the trap from linear to
quadratic dramatically changes the dependence of the entropy on
temperature, from $S\propto \ln{T^{9/2}}$ at high $T$ and
$B^\prime$, to $S\propto \ln{T^{3}}$ at low $T$ and $B^\prime$ (see
Fig.~\ref{fig:V0andS}).  The crossover temperature, $T_{\mathrm x}$,
between these two regimes occurs roughly when the high and low
temperature approximations for $V_0(T)$ (Eqns.~\ref{eq:V0high}
and~\ref{eq:V0low}) are equal. The steep nature of $S(T)$ near
$T_{\mathrm x}$ for $\mu B^\prime \simeq m g$ indicates that
adiabatic expansion from a wide range of initial temperatures leads
to a roughly constant final $T\simeq T_{\mathrm x}$, which is about
$0.1 U_0/k_B$.  The steep part of $S(T)$ near  $T_{\mathrm x}$ also
corresponds to the region over which the atoms are transferred into
the dipole trap. Above the crossover region, a negligible fraction
of atoms are in the dipole trap, but below this region, nearly all
the atoms are in the dipole trap.

The adiabatic expansion and transfer is similar to forced
evaporation, except that here  the ``evaporated" atoms during the
expansion are contained in the low density tails of the weakly
confining quadrupole trap. Many considerations are similar to
evaporation, such as the final temperature being set by the dipole
trap ``depth" for temperatures $T_{\mathrm x} \simeq 0.1 U_0/k_B$.
The ratio $\simeq 0.1$ is insensitive to $U_0$ and $w_0$, and is
approximately a constant for 10~$\mu$K $<U_{0}<$ 45~$\mu$K with
$w_0$=65~$\mu$m, and for 65~$\mu$m$<w_{0}<$ 190~$\mu$m with
$U_0$=45~$\mu$K. As with evaporation, there is in principle no
upper bound on the efficiency of the process~\cite{ketterle96b}, so
that nearly all the atoms could be cooled in the dipole trap from
any initial temperature. This would take nearly infinite time,
however, and in practice the actual process depends on the details
of the loss mechanisms and the elastic collision rates throughout
the trap. Therefore, the loss will be more pronounced for a trap at
lower densities with lower collision rates.

\begin{figure}[t]
\begin{center}
\includegraphics[width=3.5in]{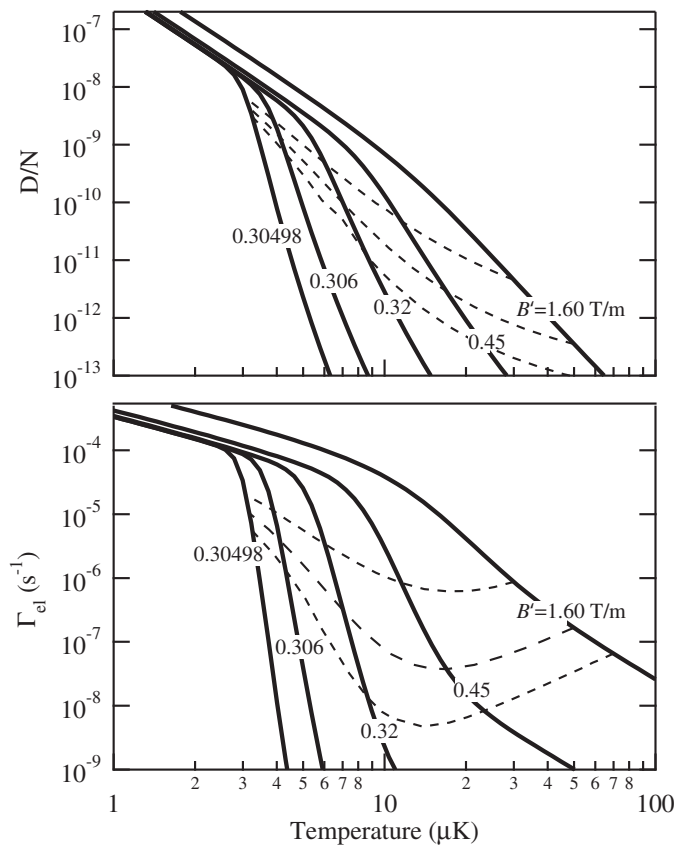}
\end{center}
\caption{{\em Top:} The phase space density per particle $D(T)/N$ in
the dipole-plus-quadrupole trap for the same field gradients as in
Fig.~\ref{fig:V0andS}, assuming no loss and thermal equilibrium
(solid lines). Example adiabatic trajectories for $D(T)/N$,
calculated at constant entropy starting at $B^\prime = 1.6$~T/m, are
shown for different starting temperatures 30, 50, and 70~$\mu$K
(dashed lines). {\em Bottom:} The average elastic collision rate,
$\Gamma_{\mathrm el}/N$ for the same field gradients (solid lines).
The collision rate is shown for the same constant entropy
trajectories as above (dashed lines).   } \label{fig:PSDandGamma}
\end{figure}

Figure~\ref{fig:PSDandGamma} shows the phase space density per
particle, $D/N$, as a function of temperature for a range of
$B^\prime$. To illustrate the expansion, several trajectories at
constant entropy are shown for a few different initial temperatures.
These plots show that if thermal equilibrium could be maintained and
absent any loss mechanism, extremely large increases in phase space
density are possible. In practice, the collision rate in the low
density tails of the atom cloud eventually drops too low to maintain
equilibrium. The actual efficiency of the process, and the optimal
trajectory, will depend on this decoupling, as well as Majorana and
one-body loss.

\section{Experiment}

The cooling process begins with a thermal beam of rubidium atoms
cooled and decelerated by a zero crossing Zeeman slower.  The slowed
atoms are loaded into a six beam MOT and then transferred to the
magnetic quadrupole trap where they are cooled by forced rf evaporation.  In the final stage of cooling the atoms are transfered
into the hybrid optical-dipole plus magnetic trap where they reach
degeneracy.

\subsection{Vacuum system}

%\begin{figure*}[tbp]
%\begin{center}
%\includegraphics[width=6.5in]{epsfiles/ExperimentalChamber}
%\end{center}
%\caption{Schematic diagram of the experiment.  From left to right the diagram shows the rubidium oven, the Zeeman slower and the experimental chamber.  Also indicated are (A) ion pumps and (B) pneumatic gate valve.}
%\label{fig:Vacuum}
%\end{figure*}

%To efficiently produce BECs it is important to understand and mitigate unwanted loss processes.  One significant source of loss is inelastic scattering with background atoms due to imperfect vacuum.  For the efficient evaporation to BEC this scattering time should be much longer than the duration of the experiment.  While our $T<20\second$ cycle time is short compared with the $\sim60\second$ evaporation time typical of magnetic trap based experiments, it is still worthwhile to realize good vacuum.

Our apparatus consists of two ultra-high vacuum (UHV) zones:
%(Fig.~\ref{fig:Vacuum})
a rubidium oven and the main experimental chamber. Each chamber is
pumped with a single $55\liter\second^{-1}$ ion pump. In addition,
the experimental chamber is pumped with a Titanium sublimation
(TiSub) pump attached to a bellows. When operated, the TiSub
filament can be positioned directly in-line with Zeeman slower,
allowing it to coat the inside of the main vacuum chamber while
avoiding the optical windows of the chamber.  After evaporation it
is retracted from the slower axis.

The oven is separated from the main experimental chamber by an
in-vacuum shutter, a differential pumping tube ($1.1\cm$ inner
diameter, $7\cm$ long), a pneumatic gate valve (normally open), and
the Zeeman slower (a $0.9\meter$ long stainless tube with a $3.5
\cm$ inner diameter).  Together these tubes give a calculated
conductance of $1.3\liter\second^{-1}$~\cite{OHanlon2003}. The
differential pumping between the chamber and the oven allow the
chamber pressure to be significantly below the oven pressure. The
oven has a nominal pressure of $5\times10^{-10}\mbar$ measured using
the ion pump current. The ion pump current in the experimental
chamber is zero to the accuracy of the pump controller (we observe a
$45\second$ magnetic trap lifetime for $T=170\uK$ atoms, consistent
with loss dominated by Majorana transitions, see
Fig.~\ref{fig:QuadrupoleLifetime}).

\begin{figure}[tbp]
\begin{center}
\includegraphics[width=3.375in]{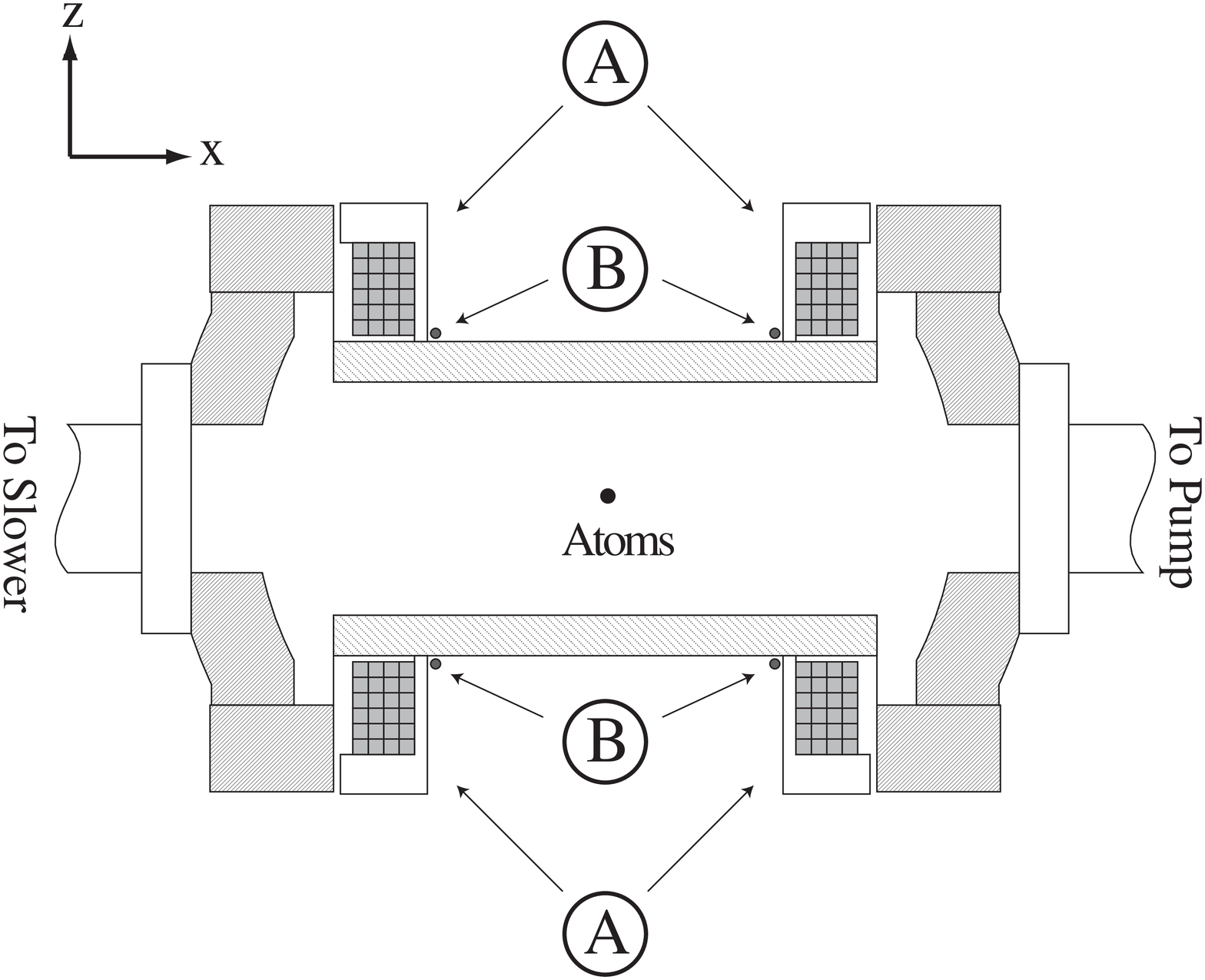}
\end{center}
\caption{Cut-away diagram of the main chamber.  The main chamber is
equipped with recessed windows in which we have mounted 24-turn
water cooled coils (A), and single turn rf evaporation loops (B). }
\label{fig:Chamber}
\end{figure}

The main chamber consists of a single stainless steel assembly with
two $14\cm$ diameter recessed windows in the top and bottom
 (Fig.~\ref{fig:Chamber}). The windows are recessed as much as
possible without blocking the line-of-sight from the remaining 16
mini-conflat ($d = 3.8\cm$) and 6 larger conflat ($d = 7\cm$)
viewports.

\subsection{Oven}

\begin{figure}[tbp]
\begin{center}
\includegraphics[width=3.375in]{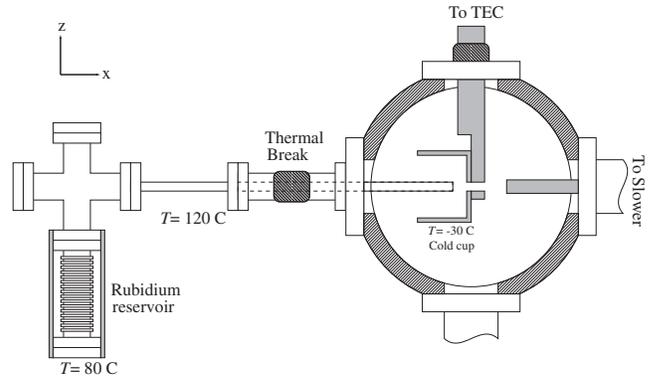}
\end{center}
\caption{Schematic diagram of rubidium oven.  The left portion of
the figure shows the heated rubidium oven, indicating the regions of
different temperature.  The right part of the figure is a cut-away
of the main oven chamber showing the end of the collimation tube and
the in-vacuum cold cup. } \label{fig:Oven}
\end{figure}

The rubidium atomic beam originates from a heated rubidium reservoir
(Fig.~\ref{fig:Oven}) and is collimated before entering the Zeeman
slower.  Efficient collimation is desirable, since it lengthens the
time between rubidium reloads, and extends the lifetime of the ion
pump.

The rubidium reservoir is contained within a $7.5\cm$ long stainless
steel (SS) bellows and was initially loaded with a $5\g$ rubidium
glass ampule.  The ampule is broken under vacuum after baking.
During operation the reservoir is held at $80\C$, setting the
rubidium vapor pressure $\approx 6\times10^{-5}\mbar$ and the mean
free path (MFP) to be $\lesssim 1\meter$ (molecular flow regime
everywhere).

The reservoir is connected to a 4-way mini-ConFlat style cross and
then to the collimating tube. The SS collimation tube has a $6\mm$
diameter and is about $22.5\cm$ in length. The 4-way cross and the
first $7.5\cm$ of the collimation tube are maintained at $120\C$,
providing a thermal gradient between the reservoir and the collimation
 tube.  The remaining $15\cm$ experiences a thermal gradient due to
 the balance of thermal conduction and black-body radiation; we estimate the temperature of the free
end of the tube to be $70\C$ (all points in the tube are above
rubidium's $39\C$ melting point).   The oven chamber is thermally
isolated from the $120\C$ oven by a glass thermal break. This oven
is not a recirculating
design~\cite{Hau1994,Walkiewicz2000,Pailloux2007}, but we expect
that the elevated temperature of the collimation tube with respect
to the reservoir keeps the tube free of rubidium.

%The oven is heated using two sections of ThermoCoax wound on the outside of the oven. One heater covers the rubidium reservoir region; the other heats the SS tube and the four-way cross.  The temperature of both regions are independently stabilized at the desired set points.  The oven is insulated first with several layers aluminum foil and then with a metalized plastic film.

We estimate that the collimation by the tube provides about a
$25\times$ improvement in flux directed into our MOT compared to a
simple aperture with the same total flux~\cite{Beijerinck1975}.  We
estimate that 1\% of all atoms departing the oven fall within the
$\approx 1\cm$ radius of the MOT beams (neglecting the transverse
heating and expansion of the beam from the Zeeman slowing process).

% clearly, if rubidium were scarce considerable improvement is possible with further collimation~\cite{Slowe2005}.

The outgoing atomic beam first passes through a $6\mm$ aperture in a
copper cold cup; after a further $2\cm$ the beam proceeds through an
in-vacuum shutter (not pictured), through a differential pumping
tube and then enters the Zeeman slower.  The cold cup is chilled to
$\approx-30\C$ by a commercial thermoelectric (TEC) based CPU cooler
and the thermal link into the vacuum chamber is a high current
vacuum feed thru. We find that the cold cup capturing excess
rubidium is essential to the long-term operation of our ion pumps,
which can fail due to excess leakage current as they pump rubidium
(similar alkali poisoning has been observed
elsewhere~\cite{RolstonPrivate}).  Even for the collimated beam,
nearly half of the outgoing rubidium is incident on the cold-cup.

To further inoculate our oven ion pump from rubidium poisoning, we
permanently maintain the pump at $70\C$ to drive off rubidium.  As
of this writing, the oven  with its initial $5\g$ rubidium sample
has been under vacuum and operating essentially daily since Feb.
2006, with an operating pressure of $5\times10^{-10}\mbar$.

\subsection{Zeeman slower}

\begin{figure}[tbp]
\begin{center}
\includegraphics[width=3.375in]{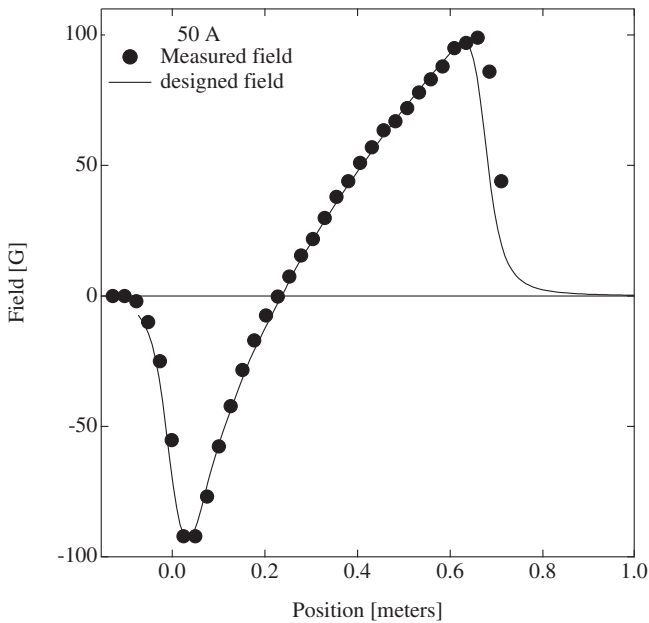}
\end{center}
\caption{Comparison of the calculated (continuous curve) and
measured field profile (points) for our Zeeman slower.  The
measurement was performed with $50\amp$ in both the forward and
reverse portions of the slower which would result in slowing at 50\%
of the peak acceleration. (We found the optimum operating condition
to be $72\amp$ in the positive field portion and $42\amp$ in the
negative field region.) } \label{fig:SlowerProfile}
\end{figure}

We slow and cool the collimated atomic beam using a Zeeman
slower~\cite{Phillips1982}.  Operating at maximum efficiency and
with an optimal field profile, our $69\cm$ long slower would stop
rubidium atoms with a peak velocity $v_{\rm
max}=390\meter\second^{-1}$ losing about $67\times10^{3}$ photon
recoil momenta ($\lambda = 780\nm$).  Our realized slower operates
at nearly $70\%$ of the maximum deceleration.

The slower field profile is a zero crossing design~\cite{witte92a},
 and is generated from two single-layer, helically wound coils which provide
the positive and negative sections of the field profile. We modeled
the field profile for the full variable-pitch coils and found an
optimized winding pattern with 84 total turns (54 and 30 turns for
the positive and negative field region, respectively). Ideally, the
optimized designed profile would slow atoms up to $84\%$ of the
maximum acceleration.  The coils were wound from electrically
insulated $1/8"$ ($3\mm$) diameter copper refrigerator tubing onto a
$d=7.6$~cm aluminum form enclosing the ConFlat vacuum tube. In
operation, cooling water forced through the copper tubes removes the
heat dissipated by the $72\amp$  and $42\amp$ flowing through the
positive and negative field portions of the slower, respectively.

The calculated and measured field profiles are shown in
Fig.~\ref{fig:SlowerProfile}.  (An additional compensation coil,
located beyond the end of the slower, zeros the field in the region
of the MOT is not shown.)  While a multi-layer design could more
optimally match the design field profile~\cite{Dedman2004}, we opted
for the simplicity afforded with a single layer coil.

The Zeeman shifted atoms are slowed by a $30\mW$ laser beam
counter-propagating with respect to the atomic beam (the slower
laser has $\sigma^+$ polarization referenced to the positive field
region of the slower).  This laser is detuned $162\MHz$ below the
$\ket{5{\rm S}_{1/2},F=2,m_F=2}$ to $\ket{ 5{\rm
P}_{3/2},F=3,m_F=3}$ $\Rb87$ cycling transition. It enters the
Zeeman slower with a $1\cm$ $1/e^2$ radius and reaches a focus in
the oven collimation tube.  A $15\mW$ repump laser beam,
copropagating with the slower laser, prepares and maintains the
atoms in the cycling transition.  This $\sigma^-$ repump beam is
detuned $165\MHz$ below the $\ket{5{\rm S}_{1/2},F=1}$ to
$\ket{5{\rm P}_{3/2},F=2}$ transition, and is resonant with the
slowed atoms in the low field portions of the slower.  We find that
the performance of the slower is fairly insensitive to
$\approx50\MHz$ changes in the repump detuning.

The rubidium beam entering the slower is at $T\approx400\kelvin$ set
by the $120\C$ portions of the oven.  Assuming a Maxwell-Boltzman
distribution the most probable velocity is
$v=340\meter\second^{-1}$.  Operating at about 70\% of the maximum
acceleration for rubidium, the peak capture velocity is
$280\meter\second^{-1}$, indicating we can slow 25\% of $\Rb87$
atoms in the full distribution.

\subsection{Magneto-optical trap}

We capture our Zeeman slowed beam of atoms in a six beam MOT at the
center of the main chamber, $15\cm$ from the end of the slower. The
required quadrupole magnetic field is generated from the same coils
used for magnetic trapping.  For the MOT we run $25\amp$
in the quadrupole coils, giving a field gradient of
$\simeq0.12\tesla\meter^{-1}$ along $\hat z$.

Each of the six MOT beam has about $20\mW$ of power and are detuned
by $-21.5\MHz$ ($3.5$ linewidths) from the cycling transition. The
MOT repumper has a total of $16\mW$ of power and is detuned by
$-5\MHz$ from the $\ket{F=1}$ to $\ket{F=2}$ transition.  The MOT
cooling and repumping light is combined in fiber, and each of the
six beams are separately collimated. The six collimated MOT beams
are directed into the vacuum chamber by reflecting them from
moveable flipper mirrors, which can be moved out of the optical
path after the atoms have been transferred to the magnetic trap and
the MOT light is no longer needed. This provides optical access for
beams needed in the experiment once the BEC has been made.

The MOT loads to $N=9\times10^8$ atoms in $3\second$, and saturates
to about $2.7\times10^9$ atoms in $20\second$.  We infer the atom
number by collecting light scattered from the MOT.  (When the cloud
is optically thick this can under-estimate the total number of atoms.)
%We measured the MOT temperature to be $()\uK$ using the MOT recapture
%technique~\cite{Chu1985}.
After $3\second$ of MOT loading, we perform polarization gradient
cooling by turning off the magnetic trap, decreasing the intensity
of the repump power to $100\uW$, and linearly increasing the
detuning of the MOT beams to 114 MHz ($19$ linewidths) over the
course of $19\ms$. At the end of the ramp the cloud has a
temperature of $29(6)\uK$~\footnote{Uncertainties reflect the
uncorrelated combination of 1-$\sigma$ statistical and systematic
uncertainties.}. We then completely extinguish the repump
light, depumping the atoms into the $F=1$ manifold.

\subsection{Magnetic Trap}\label{sect:MagTrap}

Once the atoms are laser cooled, we transfer them to the hybrid of
magnetic quadrupole and optical dipole trap. The quadrupole trap is
constructed from a pair of $24$-turn coils placed within recessed
windows above and below the main vacuum chamber
(Fig.~\ref{fig:Chamber}).  The coils are wound from Kapton insulated
$3/16"$ ($0.48\cm$) hollow square copper tube.  We flow water
through the coils in parallel with a pressure differential of about
$14~{\rm bar}$ ($200\ {\rm PSI}$) to remove the $\approx4.8\kW$
dissipated in the coils at a full current of $320\amp$.  The coils
can be effectively modeled as two 24 turn loops separated by $12\cm$
and each with a $5.8\cm$ radius (This is not the ideal
anti-Helmholtz configuration that could be realized by further
recessing our top and bottom windows. Such recessing would have
obscured optical access).  We measured a gradient of $4.8
\mT\meter^{-1}\amp^{-1}$ in the stiff direction (vertical).

We capture our laser cooled atoms by abruptly
turning on the magnetic trap. For our MOT parameters, the optimal field
gradient for best BEC production was $0.48\tesla\meter^{-1}$ along
$\hat z$. Compressing  to $1.6\tesla\meter^{-1}$ in $0.1\second$
increases the collisional rate and heats the atoms to $190(35)\uK$.
Simultaneously with the adiabatic compression, we turn on a dipole
beam along $\hat y$ to increase the density and therefore the
evaporation speed. The calculated peak collision rate and phase
space density are $15\second^{-1}$ and $5\times10^{-7}$. We then
turn on the rf, and apply a linear sweep of the frequency from
$20\MHz$ down to $3.75\MHz$ in $2.9\second$. During this rf-forced
evaporation, the temperature and number decrease to $30(5)\uK$ and
$N=8\times10^7$, and the calculated density increases to
$4.5\times10^{12}\cm^{-3}$.

\begin{figure}[tbp]
\begin{center}
\includegraphics[width=3.375in]{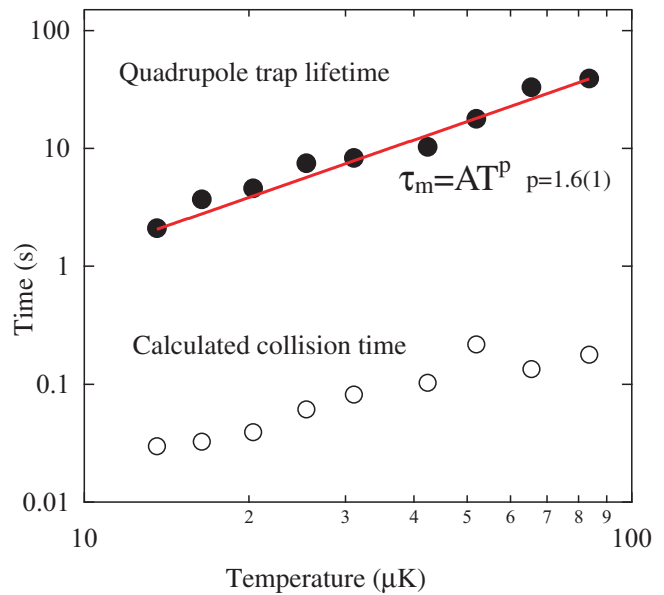}
\end{center}
\caption{Lifetime $\tau_m$ and collision times $\Gamma_{\rm{el}}^{-1}$
in the compressed quadrupole trap at the field gradient $B^{'}=1.55\tesla\meter^{-1}$.
Solid circles: measured lifetime in the quadrupole trap showing
Majorana losses.  Empty circles: calculated collision time from measured number and trap
parameters.} \label{fig:QuadrupoleLifetime}
\end{figure}

To characterize the loss from Majorana spin flips, we measure the
quadrupole trap lifetime $\tau_m$ versus the temperature $T$ at
$B^{'}=1.55\tesla\meter^{-1}$ absent the dipole trap (Fig.~\ref{fig:QuadrupoleLifetime}).
 During the lifetime measurement at each $T$, we apply a rf knife to maintain a constant $T$ given
the heating from Majorana spin flips. At our densities the calculated collision
time is approximately $200$ times shorter than $\tau_m$, which
ensures thermal equilibrium. $\tau_m$ is found to scale as
$\tau_m=AT^{p}$ with the best fit of $p$=1.6(1). A fit with fixed
$p=2.0$ yields $A=5.8\times10^{-3} \second\uK^{-2}$, approximately a
factor of 2 of that given by Eqn.~\ref{eq:Majorana}, which was
calibrated at a $\approx 50\%$ larger field gradient, $B^{'}=2.4$~$\tesla\meter^{-1}$.

\begin{figure}[tbp]
\begin{center}
\includegraphics[width=3.2in]{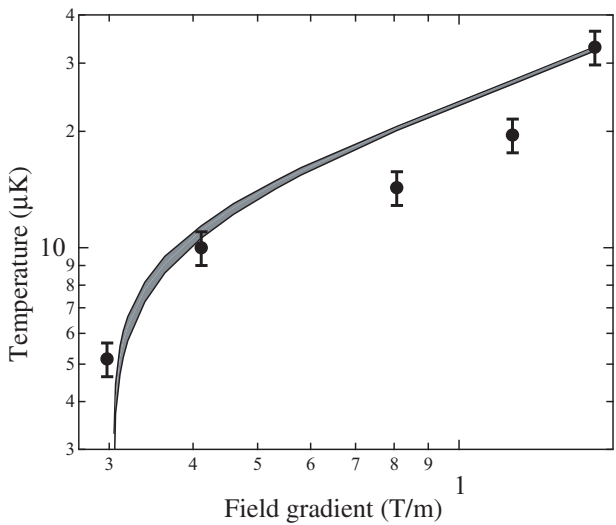}
\end{center}
\caption{Temperature $T$ versus quadrupole field gradient $B^{'}$
during the transfer into the dipole trap. Circles indicate the
measurement, where the data point at the lowest $T$ corresponds to
the end of the transfer at $\mu B^{'}<mg$, and thus free evaporation.
For comparison, the shaded region denotes calculation for a constant
entropy per atom, with the uncertainty from dipole trap parameters.} \label{fig:tempgrad}
\end{figure}

\begin{figure*}[tbp]
\begin{center}
\includegraphics[width=6.27in]{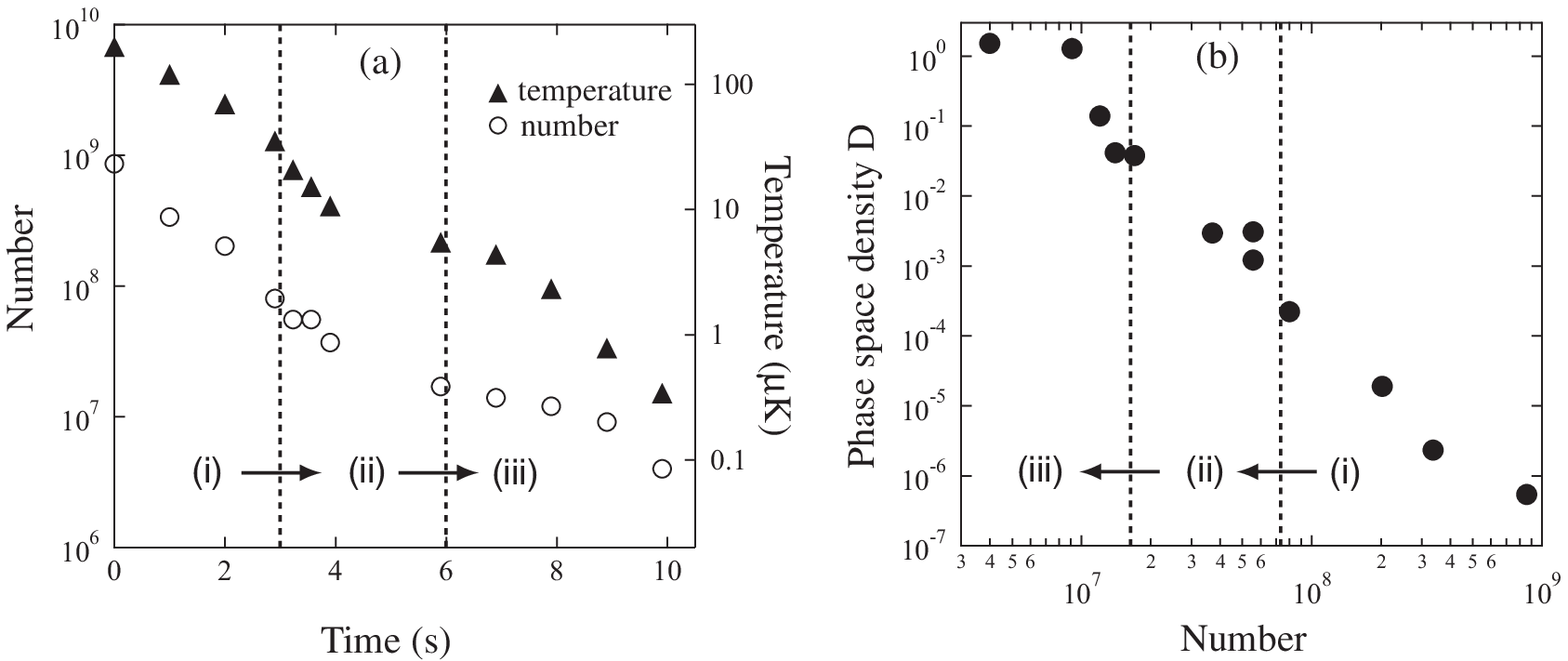}
\end{center}
\caption{Rapid production of a BEC of $N=2\times 10^6$ atoms from step (i) to (iii).
 (i) rf evaporation in the compressed
quadrupole trap (ii) loading into the dipole trap (iii) evaporation
in the dipole trap. (a) Atom number $N$ (open circles) and
temperature $T$ (solid triangles) versus the trapping time. (b)
Peak phase space density $D$ versus $N$.} \label{fig:evap}
\end{figure*}

\subsection{Transfer to the dipole trap}

Next, we adiabatically ramp down the quadrupole field gradient $B^{'}$ from
$1.6\tesla\meter^{-1}$ to $0.30\tesla\meter^{-1}$ in
$3\second$, leaving the rf power on and ramping down the frequency
to 2 MHz withing the first $1\second$, after which it remains constant.
 We turn off the rf at the end of the magnetic decompression when $B^{'}$ is smaller than the
gravity, $m g /\mu = 0.305\tesla\meter^{-1}$, and therefore loading
 the atoms into the dipole trap. The $3.5\watt$, $\lambda=1550\nm$
 dipole beam has a waist $w_0 = 65 \micron$.  We offset the center
 of dipole beam by $z_0 \sim w_0$ from the zero of
magnetic field in order to decrease the Majorana loss, where the
reduction factor scales as the density ratio $n(0)/n({\bf r}_{\rm
min})=\exp{(-E^{'}/k_B T)}$. $n(0)$ and $n({\bf r}_{\rm min})$ are the
densities at zero field point and at the trap center ${\bf r}_{\rm min}$, respectively, and
$E^{'}$ is the energy difference between ${\bf r}=0$ and ${\bf r}_{\rm min}$. We transfer the
atoms into the dipole trap with a depth $U= 49(6)\uK$ and $E^{'}=36(5)\uK$. The trap frequencies are $270\Hz$ and
$320(30)\Hz$ along $\hat x$ and $\hat z$, respectively, and $30\Hz$ along
$\hat{y}$ which is provided by the magnetic confinement at a field
of $20\uT$ along $\hat z$ at the trap bottom.

%\begin{figure}[tbp]
%\begin{center}
%\includegraphics[width=3.0in]{epsfiles/bec}
%\end{center}
%\caption{An absorption image of a Bose-Einstein condensate after a
%30 ms TOF expansion. The condensate has $N=1.8\times 10^6$
%atoms without discernable thermal components.} \label{fig:bec}
%\end{figure}

%(b) Cross-sections of
%(a) at the condensate center, along $x$ (solid squares) and the dipole
%beam direction, $y$ (open circles), respectively. Solid lines show 1D fits
%to a Thomas-Fermi profile.

During the $3\second$ loading to the dipole trap, the truncation
parameter $\eta$ increases from 6 to 10, arising from the rf knife
and $\approx 0.2\second$ of free evaporation at the end. The atom
number decreases by a factor of $\approx 4$.
Figure~\ref{fig:tempgrad} shows the temperature $T$ versus the
quadrupole field gradient $B^{'}$ during the loading. The entropy
per atom $S/Nk_{B}$ is initially 31.5, then slightly decreases by
$\lesssim 1.5$, and the cloud cools from $T=30\uK$ to $5.2\uK$.

After loading the atoms into the dipole trap, we then perform forced
evaporation by lowering the dipole trap depth from $49(6)\uK$ to a
variable $U$ in $4\second$, and simultaneously decrease $B^{'}$ from
$0.30\tesla\meter^{-1}$ to $0.27\tesla\meter^{-1}$.  The trap depth
versus time is approximately quadratic, slower toward the end the
evaporation. At $U = 4.7(7)\uK$, a bimodal distribution in the
time-of-flight (TOF) image shows a $6\ \%$ condensation fraction of the
total number $N=4\times 10^6$ with a temperature $T=0.32(5)\uK$
(calculated $T_c = 0.5\uK$). At $U=1.4(2)\uK$, the condensate is
nearly pure with $N= 1.8 \times 10^6$.
Figure~\ref{fig:evap} shows the atom number $N$ and the temperature
$T$ versus the trapping time, and the evolution of phase space
density $D$ versus $N$ in the following steps: (i) rf
evaporation in the magnetic trap (ii) transfer to the dipole
trap (iii) evaporation in the dipole trap.

We have also successfully produced condensates with similar atom numbers
adopting the same hybrid technique using different of dipole beams parameters. They
include: (1) a different waist $w_0\sim 90\micron$ still at $\lambda= 1550$ nm, and (2) a
different wavelength $\lambda= 1064$ nm and waist $w_0\sim 70\micron$. The robustness of
our approach to these changes illustrates the generality and versatility
of such hybrid magnetic and optical trapping approach.

\section{Conclusion}

We have described a simple and effective technique of producing
$\Rb87$ Bose-Einstein condensates which merges the best aspects of
magnetic and optical trapping. A condensate of $\sim 2\times 10^6$ atoms
is produced in a cycle time of $16\second$. This approach applies for a wide range of trap
 parameters, such as the quadrupole field gradient, the dipole beam trap
 depth and the waist. We expect that this technique can be generally
 applicable to other magnetically trappable atomic species.

This work was partially supported by ONR, ODNI, ARO with funds
from the DARPA OLE program, and the NSF through the JQI Physics Frontier
Center. R.L.C. acknowledges supports from NIST/NRC.

%\bibliography{main}

\end{document}